\documentclass[twocolumn,aps,prl,superscriptaddress,footinbib]{revtex4-2}
\usepackage{amsfonts}
\usepackage{amsmath}
\usepackage{amssymb}
\usepackage{color}
\usepackage{graphicx}
\usepackage{bm}
\usepackage{xcolor}

\usepackage{soul}
\usepackage{cancel}
\usepackage[colorlinks,bookmarks=false,allcolors=blue,hyperfootnotes=true]{hyperref}
\usepackage{float}				
\usepackage{multirow}

\begin{document}

\title{Spin susceptibility of nonunitary spin-triplet superconductors}

\date{\today}

\author{Thomas Bernat}
\author{Julia S. Meyer}
\author{Manuel Houzet}
\affiliation{Univ. Grenoble Alpes, CEA, Grenoble INP, IRIG, PHELIQS, 38000 Grenoble,
France}

\begin{abstract}
The spin susceptibility is an important probe to characterize the symmetry of the order parameter in unconventional superconductors. Among them, nonunitary triplet superconductors have attracted a lot of attention recently in the context of the search for topological superconductivity. Here, we derive a general formula for the spin susceptibility of nonunitary triplet superconductors within a single-band model of non-magnetic, centrosymmetric materials with strong spin-orbit coupling. We use it to critically assess experimental claims of nonunitary triplet superconductivity in some materials. 
\end{abstract}

\maketitle

{\it Introduction.--}
Unconventional superconductors exhibit exotic properties related to the unusual symmetries of the complex order parameter that describes the Cooper pair wave function~\cite{Leggetttheoreticaldescriptionnew1975,SigristPhenomenologicaltheoryunconventional1991, MineevIntroductionUnconventionalSuperconductivity1999}. In centrosymmetric crystals, pairs have either a spin-singlet or spin-triplet wave function. A further distinction exists between unitary and nonunitary spin-triplet states. In the nonunitary case, the contributions from opposite spins are unequal. As a consequence, time-reversal symmetry is broken and the excitation spectrum consists of two non-degenerate bands. Inversely, chiral triplet superconductors \cite{KallinChiralsuperconductors2016}, in which Cooper pairs bear an orbital magnetic moment, belong to the class of topological superconductors with broken time-reversal symmetry \cite{SatoTopologicalsuperconductorsreview2017} and are generically nonunitary {in the presence of spin-orbit coupling (SOC)}~\cite{YipSuperconductingstatesreduced1993}.

The spin susceptibility is a common probe of the order parameter symmetry, and can be inferred from Knight shift {or polarized neutron scattering} measurements. While it allows for a clear distinction between spin-singlet and unitary spin-triplet states, surprisingly, so far the case of nonunitary states has not been fully explored. Here we provide a theory of the {linear} spin susceptibility of nonunitary spin-triplet states and apply it to group theoretically admissible nonunitary phases of specific crystal symmetries in the presence of strong {SOC}. 

{So far, nonunitary pairing was only firmly established in the field-induced $A_1$ phase of Helium 3~\cite{Leggetttheoreticaldescriptionnew1975, wheatleyExperimentalPropertiesSuperfluid1975}. It is also strongly suspected to occur in ferromagnetic superconductors UGe$_2$, UCoGe, and URhGe because of the large mismatch of the spin bands, which favors parallel-spin pairing \cite{MineevSuperconductivityuraniumferromagnets2017}. Nonunitary triplet superconducting} phases have also been discussed in various non-magnetic materials, such as U$_{1-x}$Th$_{x}$Be$_{13}$ \cite{StewartUBe13U1xThxBe13Unconventional2019}, UPt$_3$ \cite{Joyntsuperconductingphases$UPt_3$2002}, Sr$_2$RuO$_4$ \cite{Mackenziesuperconductivity$Sr_2RuO_4$physics2003}, and UTe$_2$ \cite{AokiUnconventionalSuperconductivityUTe22022}. Our theory provides a unifying frame to interpret the spin properties of any triplet superconductor. We discuss spin susceptibility measurements in the above-mentioned non-magnetic compounds in the light of our results.

{\it Model.--} We consider a single-band model of a centrosymmetric and non-magnetic metal, such that one can distinguish between singlet and triplet superconductivity. Within the quasiclassical theory of superconductivity~\cite{EilenbergerTransformationGorkovsequation1968, LarkinQuasiclassicalMethodTheory1969, SereneQuasiclassicalApproachSuperfluid1983, kopninTheoryNonequilibriumSuperconductivity2001}, the components of the magnetization carried by the spins of the conduction electrons, $\mathbf{M}=(M_x,M_y,M_z)$, are~\cite{BergeretOddTripletSuperconductivity2005}
\begin{equation}
\label{eq:mag}
M_a=M^N_{a}+\frac i4\pi \nu_0 \mu_B  g_a T\sum_{\omega}\mathrm{Tr}\,\langle  \sigma_a \tau_z \check g_{\mathbf{k},\omega}\rangle_{\mathbf{k}}
\end{equation}
($a=x,y,z$). Here, the first term is the normal-state contribution, $M^N_{a}=\chi^N_a H_a$,  which is induced by an external magnetic field $\mathbf{H}=(H_x,H_y,H_z)$ with components $H_a$ along the principal axes of the magnetic susceptibility tensor. It is determined by $\chi_a^N=(g_a^2/4)\chi_P$, where $\chi_P=2\nu_0\mu_B^2$ is the Pauli susceptibility, $\nu_0$ is the density of states per spin in the normal state, $\mu_B$ is the Bohr magneton, and $g_a$ are (possibly anisotropic) Land\'{e} factors~\footnote{The assumption of centrosymmetric Bloch bands guarantees the presence of Kramers pairs at fixed $\mathbf{k}$. In pseudo-spin bands, the Kramers degeneracy index behaves as a spin $\frac12$. Then, the Zeeman coupling takes a diagonal form after a suitable choice of the spin-quantization axes, cf.~Chap.~15.6 in Ref.~\cite{AbragamElectronParamagneticResonance2012}. While the sign of $g_a$ is arbitrary, their product $g_xg_yg_z$ is an invariant. We did not see signatures of that invariant in the spin susceptibility. Note also that our formalism is readily extended to the case of $\mathbf{k}$-dependent $g_a$, which may be important in some crystals, where the Kramers degeneracy does not behave as a spin $\frac 12$~\cite{SamokhinSpinsusceptibilitysuperconductors2021}.}. The second term in Eq.~\eqref{eq:mag} is induced by superconducting correlations. It is expressed in terms of the quasiclassical Green function $\check g_{\mathbf{k},\omega}$ at momentum $\mathbf{k}$ on the Fermi surface and Matsubara frequency $\omega=(2n+1)\pi T$ (integer $n$), where $T$ is the temperature. The quasiclassical Green function $\check g_{\mathbf{k},\omega}$ is a $4\times 4$ matrix in spin and Nambu spaces with associated Pauli matrices $\sigma_{x,y,z}$ and $\tau_{x,y,z}$, respectively. Furthermore, $\langle \cdots\rangle_{\mathbf{k}}$ denotes averaging over the Fermi surface and we take units with $\hbar=k_B=1$.

To evaluate Eq.~\eqref{eq:mag}, we need $\check g_{\mathbf{k},\omega}$ that solves
\begin{equation}
\label{eq:eilenberger}
[\omega\tau_z + \check{\Delta}_\mathbf{k} - i\mathbf{h}\cdot \bm{\sigma}, \check{g}_{\mathbf{k},\omega}] = 0,
\end{equation}
together with the normalization conditions $\check g_{\mathbf{k},\omega}^2=1$ and $\mathrm{Tr}\,\check g_{\mathbf{k},\omega}=0$. Equation~\eqref{eq:eilenberger} accounts for the Zeeman field, $\mathbf{h}=(h_x,h_y,h_z)$ with $h_a=\frac12g_a\mu_B H_a$, and the spin-triplet order parameter $\mathbf{d}_\mathbf{k}$ that enters the gap matrix
\begin{equation}
\check{\Delta}_\mathbf{k} = \begin{pmatrix}
0 & \hat{\Delta}_\mathbf{k} \\
\hat{\Delta}_\mathbf{k}^\dagger & 0
\end{pmatrix}\quad \mathrm{with}\quad \hat{\Delta}_\mathbf{k}=\mathbf{d}_\mathbf{k} \cdot \bm{\sigma}.
\end{equation}

According to the symmetry analysis of unconventional superconductivity~\cite{SigristPhenomenologicaltheoryunconventional1991,MineevIntroductionUnconventionalSuperconductivity1999}, a nonunitary phase may appear when the superconducting transition occurs in an irreducible representation (IR), $\Gamma$, of the crystal's point group with dimension $d_\Gamma\geq 2$.
The order parameter then factorizes as
\begin{equation}
\mathbf{d}_\mathbf{k} = \Delta(T) \sum_{\alpha=1}^{d_\Gamma} \eta_\alpha \bm{\psi}_{\mathbf{k}, \alpha}, 
\qquad \sum_{\alpha=1}^{d_\Gamma} |\eta_\alpha|^2=1,
\end{equation}
where $\Delta(T)$ is a common, $T$-dependent factor, and the $\eta_\alpha$-coefficients are weights in front of the basis functions of $\Gamma$. At strong SOC, the spin-momentum structure is described by the vectors $\bm{\psi}_{\mathbf{k},\alpha}$, such that
\begin{equation}
\langle \bm{\psi}_{\mathbf{k},\alpha} \cdot \bm{\psi}_{\mathbf{k},\beta}^{*} \rangle_\mathbf{k} = \delta_{\alpha\beta},
\end{equation}
which are locked to the same principal axes as for the magnetic response in the normal state. The set of complex coefficients $\Delta(T)\eta_\alpha$ should satisfy the self-consistent gap equations,
\begin{equation}
\Delta(T)\eta_\alpha = \dfrac{\lambda \pi T }{2} \sum_{|\omega|<{\cal E}}  \mathrm{Tr} \,
\langle \bm{\psi}_{\mathbf{k},\alpha}^{*} \cdot  \bm{\sigma} \tau_- \check{g}_{\mathbf{k},\omega} \rangle_\mathbf{k},
\label{eq:gap_eq_QCGF}
\end{equation} 
where we introduced $\tau_- = (\tau_x-i\tau_y)/2$, and $\lambda$  is the effective pairing amplitude taken constant within an energy window, $\cal E$, around the Fermi level.

{\it Susceptibility.--}
To find the linear response for the magnetization, we need to solve Eqs.~\eqref{eq:eilenberger}-\eqref{eq:gap_eq_QCGF} perturbatively in the field with $\check{g}_{\mathbf{k},\omega}=\check{g}_{\mathbf{k},\omega}^{(0)}+\check{g}_{\mathbf{k},\omega}^{(1)}+\dots$ In zeroth order in the field, we find
\begin{equation}
\check{g}_{\mathbf{k},\omega}^{(0)} = \dfrac{1}{2} \check{\Omega}_\mathbf{k} \sum_\pm \dfrac{1}{\Omega_{\mathbf{k},\pm}} \left(  1  \pm \hat{\mathbf{q}}_\mathbf{k}\cdot\bm{\sigma} \tau_z \right),
\quad \hat{\mathbf{q}}_\mathbf{k} = \frac{\mathbf{q_k}}{|\mathbf{q_k}|},
\label{eq:g0_NU}
\end{equation}
where $\check{\Omega}_\mathbf{k} = \omega \tau_z + \check{\Delta}_\mathbf{k}$ and $\Omega_{\mathbf{k},\pm} =( \omega^2 + |\mathbf{d_k}|^2 \pm |\mathbf{q_k}|)^{1/2}$ with $\mathbf{q_k} = i\mathbf{d_k}\times\mathbf{d}_\mathbf{k}^*$ {real}. 

Nonunitary phases are characterized by $\mathbf{q_k}\neq 0$. Thus their (angle-resolved) quasiparticle density of states at energy $E$, $\nu_\mathbf{k}(E)=(\nu_0/4) \mathrm{Re}\,[\mathrm{Tr}\, \tau_z\check g_{\mathbf{k},\omega}]$ after analytic continuation $i\omega\to E+i0^+$, displays a two-gap structure with gaps $\Delta_{\mathbf{k},\pm} =( |\mathbf{d}_\mathbf{k}|^2\pm|\mathbf{q_k}|)^{1/2}$. We further distinguish nonunitary phases with $\mathbf{d}_\mathbf{k}^2=0$ from the generic case, $\mathbf{d}_\mathbf{k}^2\neq 0$. In the former case, $\Delta_{\mathbf{k},-}=0$ and the gap vanishes over the whole Fermi surface~\footnote{A vanishing gap over the Fermi surface seems to be in contradiction with the Blount theorem, which states that the gap can only vanish at point nodes in triplet superconductors with strong SOC~\cite{BlountSymmetrypropertiestriplet1985}, at least in symmorphic crystals~\cite{NormanOddparityline1995}. However, one should note that the property $\mathbf{d}_\mathbf{k}^2=0$ is only approximate if one uses the most general form of admissible basis functions for $\Gamma$~\cite{YipSuperconductingstatesreduced1993}.}. It reflects that electrons with spins (anti)parallel with $\mathbf{q_k}$ are (un)paired. 

In general, to find which combination of the basis functions is favored below the critical temperature, $T_c\simeq 1.13\,{\cal E} e^{-1/\lambda}$ in the weak-coupling regime $\lambda\ll 1$, and what is the associated gap $\Delta(T)$, one should minimize the energy functional whose saddle point is given by the gap equation, Eq.~\eqref{eq:gap_eq_QCGF}. Actually nonunitary phases are seldom favored when $\lambda\ll 1$ (Refs.~\cite{KuznetsovaPairingstatemulticomponent2005,MukherjeeMicroscopictheoriescubic2006} work out some examples within Ginzburg-Landau (GL) theory, at $T_c-T\ll T_c$). The feedback of spin fluctuations on the order parameter (akin to strong coupling) is usually invoked for stabilizing them~\cite{SugiyamaMechanismStabilizeNon1995}. In this work, we assume that these effects allow fixing $T$-independent\footnote{The $T$-independence of the coefficients $\eta_\alpha$ is readily seen from the minimization of phenomenological {\color{blue} GL} theories~\cite{SigristPhenomenologicaltheoryunconventional1991}, as well as from group-theory arguments. It does not persist when two IRs with close $T_c$s are considered.} weights $\eta_\alpha$ such that $\mathbf{q_k}\neq 0$. Then the self-consistent Eqs.~\eqref{eq:gap_eq_QCGF}, together with Eq.~\eqref{eq:g0_NU}, reduce to a single equation,
\begin{equation}
\dfrac{1}{\lambda} = {\frac{\pi T}{2}}  \sum_{\omega,\pm} \left\langle \dfrac{|\bm{\psi}_\mathbf{k}|^2 \pm |\bm{\psi}_\mathbf{k} \times \bm{\psi}_\mathbf{k}^*|}{\sqrt{\omega^2 + \Delta^2(T) (|\bm{\psi}_\mathbf{k}|^2 \pm |\bm{\psi}_\mathbf{k} \times \bm{\psi}_\mathbf{k}^*|)}} \right\rangle_\mathbf{k}.
\label{eq:gap_equation}
\end{equation}
Here $\bm{\psi}_\mathbf{k} = \sum_\alpha \eta_\alpha \bm{\psi}_{\mathbf{k}, \alpha}$, such that $\mathbf{d_k} = \Delta(T) \bm{\psi}_\mathbf{k}$ and $\langle| \bm{\psi}_\mathbf{k}|^2 \rangle_\mathbf{k} = 1$.

Note that $\langle \mathbf{q_k}\rangle_\mathbf{k}$ is the spin carried by the Cooper pairs' condensate. (It may vanish even in nonunitary phases.) However, in the absence of a magnetic field, the total magnetization remains zero~\cite{Leggetttheoreticaldescriptionnew1975}, as readily shown inserting Eq.~\eqref{eq:g0_NU} into \eqref{eq:mag}~\footnote{The compensation of the Cooper pair magnetization by unpaired electrons is a feature of the quasiclassical approximation (i.e., large Fermi energy).}.  To find the contribution of superconductivity to Eq.~\eqref{eq:mag}, we thus need the first-order correction in the field,
\begin{eqnarray}
\label{eq:g1_NU}
\check{g}_{\mathbf{k},\omega}^{(1)}
&=& 
\dfrac{1}{\Omega_{\mathbf{k},+}\Omega_{\mathbf{k},-}(\Omega_{\mathbf{k},+}+\Omega_{\mathbf{k},-})} 
\left\{ -i\check{\Omega}_\mathbf{k} [\mathbf{h}\cdot\bm{\sigma} \tau_z , \check{\Delta}] \phantom{\dfrac{1}{1}} \right. 
 \\ 
&&
\left.
- \left(
\dfrac{\Omega_{\mathbf{k} ,+}-\Omega_{\mathbf{k} ,-}}{\Omega_{\mathbf{k} ,+}+\Omega_{\mathbf{k} ,-}}
\hat{\mathbf{q}}_\mathbf{k}\cdot\bm{\sigma}\tau_z + \dfrac{1}{2} \right) ( \mathbf{q_k} \times\mathbf{h}\cdot\bm{\sigma} ) 
\right\}.
\nonumber
\end{eqnarray}
Inserting $\check{g}_\mathbf{k}^{(1)}$ into the r.h.s.~of Eq.~\eqref{eq:gap_eq_QCGF} does not yield additional contributions. Thus $\Delta(T)$ does not depend on the field in linear order in $\mathbf{h}$. We then insert Eq.~\eqref{eq:g1_NU} into \eqref{eq:mag} to find the linear susceptibility, $\chi^S_{ab}=\partial M_a/\partial H_b$~\footnote{Alternative forms can be obtained using  $|\mathbf{d_\mathbf{k}}|^2\mathrm{Re}(d_{\mathbf{k},a}d_{\mathbf{k},b}^*)=\mathrm{Re}(\mathbf{d}_\mathbf{k}^{2}d_{\mathbf{k},a}^*d_{\mathbf{k},b}^*)+\frac12(\mathbf{q}_\mathbf{k}^2\delta_{ab}- q_{\mathbf{k},a} q_{\mathbf{k},b})$.}: 
\begin{widetext}
\begin{equation}
\dfrac{4 \chi^S_{ab}}{ g_a g_b \chi_P } = 
\delta_{ab} - 2\pi T \sum_{\omega}  
\left\langle  \dfrac{ {(\Omega_{\mathbf{k},+}+\Omega_{\mathbf{k},-})^2 \mathrm{Re}(\mathbf{d}_\mathbf{k}^2 d_{\mathbf{k},a}^*d_{\mathbf{k},b}^* )
+ {(\omega^2+\Omega_{\mathbf{k},+}\Omega_{\mathbf{k},-})} (\mathbf{q}_\mathbf{k}^2\delta_{ab} - q_{\mathbf{k},a} q_{\mathbf{k},b} )}}{|\mathbf{d_k}|^2\Omega_{\mathbf{k},+}\Omega_{\mathbf{k},-}(\Omega_{\mathbf{k},+}+\Omega_{\mathbf{k},-})^3}
 \right\rangle_\mathbf{k},
\label{eq:chi_NU}
\end{equation}
\end{widetext}
evaluated with $\Delta(T)$ obtained from Eq.~\eqref{eq:gap_equation}. To the best of our knowledge, this formula is new.  Reference~\cite{HiranumaParamagneticEffectsj2021} derives a formula that is equivalent to Eq.~\eqref{eq:chi_NU} for two components of the spin susceptibility tensor only in the case of the point group $D_{2h}$, which we discuss below. In the following, we analyze our main result, Eq.~\eqref{eq:chi_NU}, in different regimes.

At $\mathbf{q_k}=0$, Eq.~\eqref{eq:chi_NU} reduces to the textbook formula for unitary triplet phases~\cite{Leggetttheoreticaldescriptionnew1975,MineevIntroductionUnconventionalSuperconductivity1999}:
\begin{equation}
\dfrac{4 \chi^S_{ab}}{ g_a g_b \chi_P }  = \left\langle Y_\mathbf{k}(T) \hat{d}_{\mathbf{k},a}\hat{d}_{\mathbf{k},b}^* + \left(\delta_{ab} - \hat{d}_{\mathbf{k},a}\hat{d}_{\mathbf{k},b}^*\right) \right\rangle_\mathbf{k}
\label{eq:unitary}
\end{equation}
with $\hat{\mathbf{d}}_\mathbf{k}=\mathbf{d_k}/|\mathbf{d_k}|$ and the angle-resolved Yosida function,
\begin{equation}
Y_\mathbf{k}(T) =1-\pi T\sum_\omega\frac{|\mathbf{d_k}|^2}{(\omega^2+|\mathbf{d_k}|^2)^{3/2}},
\end{equation}
which also appears in the spin susceptibility of spin-singlet superconductors~\cite{YosidaParamagneticSusceptibilitySuperconductors1958}, replacing $\mathbf{d_k}$ with the singlet order parameter. In particular, $Y_\mathbf{k}(0)=0$ and $Y_\mathbf{k}(T_c)=1$. In the unitary case, $\mathbf{d_k}$ is the direction along which the spin projection of the Cooper pairs associated with a given $\mathbf{k}$ vanishes. Thus, as discussed in the literature, Eq.~\eqref{eq:unitary} describes that both Cooper pairs and unpaired electrons contribute to the magnetization when $\mathbf{H}\perp \mathbf{d_k}$: the susceptibility is the same as in the normal state. However, only unpaired electrons contribute to the magnetization when $\mathbf{H}\parallel \mathbf{d_k}$: the susceptibility vanishes as $T\to 0$. 

At $\mathbf{q_k}\neq 0$, Eq.~\eqref{eq:chi_NU} contains interference effects from the two bands with different gaps. Thus it cannot be reduced to an expression similar to \eqref{eq:unitary}. This is clearly seen in the nonunitary case with $\mathbf{d}_\mathbf{k}^2=0$, 
\begin{equation}
\dfrac{4 \chi^S_{ab}}{ g_a g_b \chi_P }  = 
\left\langle  
\hat{q}_{\mathbf{k},a}\hat{q}_{\mathbf{k},b} + X_\mathbf{k}(T)\left(\delta_{ab} - \hat{q}_{\mathbf{k},a}\hat{q}_{\mathbf{k},b}\right) 
\right\rangle_\mathbf{k},
\label{eq:nonunitary-d20}
\end{equation}
where 
\begin{equation}
X_\mathbf{k}(T) = 1-\pi T\sum_\omega\frac{2|{\mathbf{q_k}}|}{\sqrt{\omega^2+2|{\mathbf{q_k}} |} (|\omega|+\sqrt{\omega^2+2|{\mathbf{q_k}} |} )^{2}}.
\label{eq:NotYosida}
\end{equation}
In particular, $X_\mathbf{k}(0)=\frac12$ and $X_\mathbf{k}(T_c)=1$. As $\mathbf{q_k}$ is the Cooper pairs' spin direction associated with a given $\mathbf{k}$, Eq.~\eqref{eq:nonunitary-d20} describes that all paired and unpaired electrons contribute to the susceptibility when $\mathbf{H}\parallel \mathbf{q_k}$, as in the normal state. However, only unpaired electrons contribute to the magnetization when $\mathbf{H}\perp \mathbf{q_k}$: as $T\to 0$, the susceptibility is reduced to half of its normal-state value, which is the susceptibility of the unpaired electrons in the ``$-$'' band. In particular, if $\mathbf{q_k}$ keeps a constant direction in space, then the susceptibility tensor gets suppressed along two directions, while it keeps its normal-state value along a third one. This is in contrast with the unitary case, where the susceptibility tensor keeps its normal-state value along two directions and is suppressed along a third one, if $\mathbf{d_k}$ keeps a constant direction in space. In general, the direction of $\mathbf{q_k}$ (or $\mathbf{d_k}$) varies with $\mathbf{k}$ and the average over the Fermi surface mixes the behavior of the three directions.

Considering the generic case, we use $\mathbf{q}^2_\mathbf{k}=|\mathbf{d_k}|^4-|\mathbf{d}^2_\mathbf{k}|^2$ to find that Eq.~\eqref{eq:chi_NU} simplifies at zero temperature, 

\begin{eqnarray}
\label{eq:suscNU_T0}
&&\dfrac{4 \chi^S_{ab}}{ g_a g_b \chi_P }
	=\frac 12
	\delta_{ab} 
	+\frac 12\left\langle \hat q_{\mathbf{k},a} \hat q_{\mathbf{k},b} \right\rangle_\mathbf{k}
\\&&\quad
	-
	\left\langle  \ln\left(\frac{\Delta_{\mathbf{k},+}}{\Delta_{\mathbf{k},-}}\right)
\frac{\mathrm{Re}[
\mathbf{d}_\mathbf{k}^2(|\mathbf{d_k}|^2d^*_{\mathbf{k},a}-\mathbf{d}^{*2}_\mathbf{k}d_{\mathbf{k},a})d^*_{\mathbf{k},b}]}{|\mathbf{q}_\mathbf{k}|^3}\right\rangle_{\mathbf{k}}.
\nonumber
\end{eqnarray}
Here the first line is the same as Eq.~\eqref{eq:nonunitary-d20} with $\mathbf{d}^2_\mathbf{k}=0$ at $T=0$.
Near $T_c$, we use Eq.~\eqref{eq:gap_equation} to find
\begin{equation}
\label{eq:chi-Tc}
\dfrac{4 \chi^S_{ab}}{ g_a g_b \chi_P }
=\delta_{ab} -
\frac{{2}{\mathrm{Re} \langle  \bm{\psi}_{\mathbf{k},a}\bm{\psi}_{\mathbf{k},b}^* \rangle_\mathbf{k}}}
{\left\langle 
|\bm{\psi}_\mathbf{k}|^4+|\bm{\psi}_\mathbf{k}\times \bm{\psi}_\mathbf{k}^*|^2 \right\rangle_\mathbf{k}}\left(1-\frac T{T_c}\right).
\end{equation}
Finally, we find at any temperature that the trace of the spin susceptibility takes a rather simple form,
\begin{equation}
\sum_a \frac{4\chi^S_{aa}}{g_a^2\chi_P}
=
3 - 2\pi T \sum_\omega 
\left\langle 
\dfrac{\Omega_{\mathbf{k},+}\Omega_{\mathbf{k},-}-\omega^2}{\Omega_{\mathbf{k},+}\Omega_{\mathbf{k},-}(\Omega_{\mathbf{k},+} +\Omega_{\mathbf{k},-})} 
\right\rangle_\mathbf{k}.
\label{eq:trace}
\end{equation}
It reduces to $2 + \langle Y_\mathbf{k}(T)\rangle_\mathbf{k}$ in the unitary case~\footnote{
Ref.~\cite{Leggetttheoreticaldescriptionnew1975} stresses that Eq.~\eqref{eq:unitary} does not apply to the nonunitary case, and quotes the PhD thesis of S. Takagi, University of Tokyo (1973) -- to which we do not have access -- for its generalization. 
Our Eq.~\eqref{eq:trace} disagrees with another statement in Ref.~\cite{Leggetttheoreticaldescriptionnew1975} that the trace of Eq.~\eqref{eq:unitary} also holds for the nonunitary case at any temperature.} and $1 + 2\langle X_\mathbf{k}(T)\rangle_\mathbf{k}$ in the nonunitary case with $\mathbf{d}^2_\mathbf{k}=0$. At $T=0$ it equals 2 for any (unitary or nonunitary) triplet state.

Possible nonunitary phases have been studied experimentally in various crystals belonging to different symmetry classes.  Below we analyze the properties of the susceptibility tensor, Eq.~\eqref{eq:chi_NU}, for the nonunitary phases that can appear in cubic ($O_h$), hexagonal ($D_{6h}$), and tetragonal ($D_{4h}$) symmetry classes~\cite{SigristPhenomenologicaltheoryunconventional1991}, which are listed in Table~\ref{tab:table2} together with a selection of representative basis functions for the pairing state~\cite{SigristPhenomenologicaltheoryunconventional1991,MineevIntroductionUnconventionalSuperconductivity1999,YipSuperconductingstatesreduced1993}. We put the results in relation with the experimental findings.
\begin{table}[t]
\caption{
Possible nonunitary states and representative basis functions for the cubic ($O_h$), hexagonal ($D_{6h}$), and tetragonal ($D_{4h}$) point groups with strong SOC~\cite{SigristPhenomenologicaltheoryunconventional1991}. In $D_{6h}$ and $D_{4h}$ we allow for the superposition of two basis functions with real coefficients $c_{1}(\hat{\mathbf{k}})$ and $c_{2}(\hat{\mathbf{k}})$ that are invariant under all symmetries of the point group~\cite{YipSuperconductingstatesreduced1993}, such that the phases are unitary if $c_1=0$. Furthermore, $\varepsilon = e^{2i\pi/3}$ and $\hat k_+=\hat k_x+i\hat k_y$.
}
\label{tab:table2}
\begin{ruledtabular}
\begin{tabular}{cccc}
 \begin{tabular}{c}Point\\group \end{tabular} & IR & \begin{tabular}{c} Pairing state \\ $\bm{\psi}_\mathbf{k}\propto$ \end{tabular} & $\langle \mathbf{q_k}\rangle_{\mathbf{k}} \propto$ \\ \hline 
\multirow{5}{*}{$O_h$}	& ${E_{u}}$ & $(\hat k_x, \varepsilon \hat k_y, \varepsilon^2 \hat k_z) $ & 0 \\

                        &  $F_{1u}$  & $(\hat k_y+i\hat k_z, -\hat k_x, -i\hat k_x)$ & $(1,0,0)$ \\
                     	& $F_{1u}$ & $(\varepsilon \hat k_y - \varepsilon^2 \hat k_z, \hat k_z - \varepsilon \hat k_x, \varepsilon^2 \hat k_x - \hat k_y)$ & $(1,1,1)$ \\                     	
						& $F_{2u}$ & $(\varepsilon\hat  k_y + \varepsilon^2 \hat k_z,\hat  k_z + \varepsilon \hat k_x, \varepsilon^2 \hat k_x + \hat k_y)$ & $(1,1,1)$ \\ 
                     	& $F_{2u}$ & $(\hat k_y+i\hat k_z, \hat k_x, i\hat k_x)$ & $(1,0,0)$ \\ \hline
\multirow{2}{*}{$D_{6h}$}	& ${E_{1u}}$ & $c_1(\hat{\mathbf{k}})\hat k_z(1,i, 0)+c_2(\hat{\mathbf{k}})\hat k_+ (0, 0, 1)$ & $(0,0,1)$ \\
                     		& $E_{2u}$ & $c_1(\hat{\mathbf{k}})\hat k_+(1,i,0)+ c_2(\hat{\mathbf{k}}) \hat k_z\hat k_+^2 (0,0,1)$ & $(0,0,1)$ \\ \hline 
$D_{4h}$ & ${E_{u}}$ & $c_1(\hat{\mathbf{k}})\hat k_z(1, i, 0)+c_2(\hat{\mathbf{k}})\hat k_+ (0, 0, 1)$ & $(0,0,1)$ \\
\end{tabular}
\end{ruledtabular}
\end{table}

\begin{figure}
		\centering
		\includegraphics[]{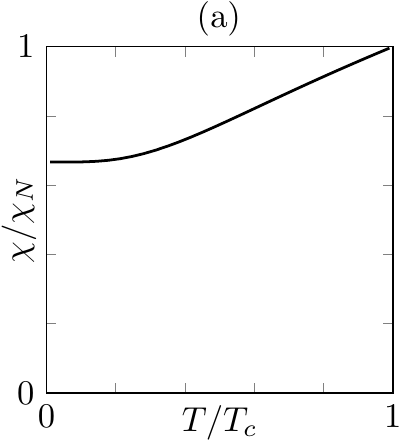}
		\hfill
		\includegraphics[]{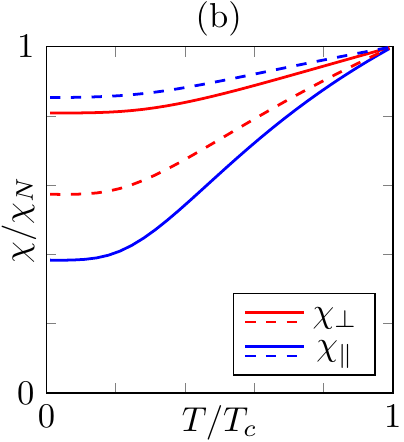}
\caption{
Temperature dependence of the principal values of the spin susceptibility tensor, in units of {$\chi_N = (g^2/4)\chi_P$}, for various {nonunitary} triplet phases in $O_h$ with pairing functions in the Table: (a) line 1, where the susceptibility is isotropic; (b) lines 2, 3, and 5 in solid and line 4 in dashed. The doubly and singly-degenerate components $\chi_\perp$ and $\chi_\parallel$ are for principal axes perpendicular and parallel to $\langle\mathbf{q_k}\rangle_\mathbf{k}$, respectively.}
\label{fig:1}
\end{figure}

{\it Cubic symmetry.--} In $O_h$, the normal-state susceptibility tensor is isotropic: $g_{x,y,z}=g$. Among possible nonunitary phases, we first consider the first line in the Table, where $\langle \mathbf{q_k}\rangle_\mathbf{k}=0$. Thus, there is no preferential direction and we find an isotropic susceptibility, cf.~Fig.~\ref{fig:1}(a). In particular, ${\chi^S_{ab}}(0)/\chi_P=(g^2/6)\delta_{ab}$ at $T=0$; the point nodes along the direction $\hat{\mathbf{k}}=1/\sqrt{3}(1,1,1)$, as well as equivalent ones, result in $\chi^S_{ab}(T)-\chi^S_{ab}(0)\propto T^2$ as $T\to 0$.

Next, we consider the second line in the Table, where $\langle \mathbf{q_k}\rangle_\mathbf{k} \propto (1, 0, 0)$. Therefore, the susceptibility develops an anisotropy along the direction $\hat x$. 
The $T$-dependence of the susceptibility tensor's principal values is shown in {solid lines} in {Fig.~\ref{fig:1}(b)}. At low temperature, the deviation from the $T=0$ result is again proportional to $T^2$ because of point nodes at $k_y=k_z=0$. The {third and} fifth lines of the Table give the same result, with different principal axes determined by $\langle \mathbf{q_k} \rangle_\mathbf{k}$.
The fourth line in the Table yields a different temperature dependence, shown in dashed line in {Fig.~\ref{fig:1}(b)}. Note that the hard/easy axes are reversed compared to the three previous cases.

The symmetry group $O_h$ applies to the cubic crystal U$_{1-x}$Th$_x$Be$_{13}$, which exhibits two superconducting phases at finite doping ratio $x$. The lower one in temperature was suspected to be nonunitary \cite{SigristPhenomenologicaltheorysuperconductivity1989}. However, since Ref.~\cite{Sonier$$KnightShift2000} {reported} a constant Knight shift in at least one direction, it does not match the reduction with temperature that we found in all considered nonunitary phases.

\begin{figure}
	\centering
	\includegraphics[]{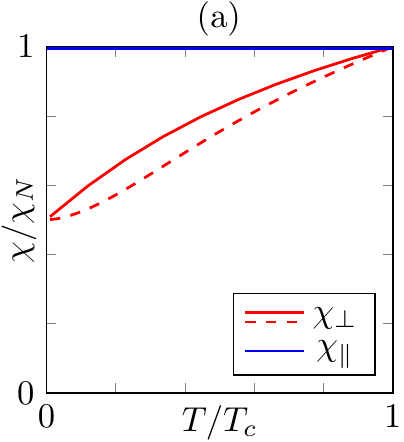}
	\hfill
	\includegraphics[]{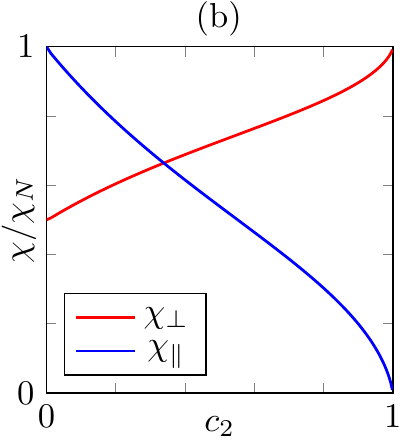}
	\caption{
	(a) Temperature dependence of the principal values of the spin susceptibility tensor, in units of {$\chi_N = (g^2/4)\chi_P$, assumed isotropic for simplicity}, for nonunitary triplet pairing states in 
	$D_{6h}$ and $D_{4h}$. 
	The first line in the Table for $D_{6h}$ and $D_{4h}$ yield the same result shown in solid, whereas the second line for $D_{6h}$ is shown in dashed, all with $c_2=0$. Here, $\chi_\perp$ and $\chi_\parallel$ are for the principal axes perpendicular and parallel to $\mathbf{q}_\mathbf{k}\propto \hat z$. 
	(b) Dependence on {$c_2$} of the principal values of the spin susceptibility tensor at $T=0$ for the triplet pairing state $\sqrt{3/2}[c_1\hat k_z(1,i,0)+c_2\hat k_+(0,0,1)]$ with $c_1^2+c^2_2=1$ (solid lines in Fig.~2(a)). At $c_2=1$, the phase is unitary, yielding the characteristic vanishing of one component of the susceptibility tensor, while the other two retain their normal-state value.}
	\label{fig:2}
\end{figure}

{\it Hexagonal symmetry.--} Assuming $c_2=0$ in the Table for $D_{6h}$, we find two examples of nonunitary phases with $\mathbf{d}^2_\mathbf{k}=0$ and $\mathbf{q_k}\propto (0,0,1)$. The susceptibility tensor's principal values are then determined by Eq.~\eqref{eq:NotYosida}. The presence of a line node at $k_z=0$ and point node at $k_x=k_y=0$ yield dependences $\propto T$ and $T^2$, respectively, at $T\to 0$ for the deviation of the two principal values that are affected by superconductivity with respect to their $T=0$ value, cf.~Fig.~\ref{fig:2}(a). However, we recall that the property $\mathbf{d}^2_\mathbf{k}=0$ is fragile when more general basis functions are considered. In particular, it is lost for nonunitary phases with $c_2\neq 0$. In {Fig.~\ref{fig:2}(b)}, we plot the dependence of the principal values $\chi_{xx}=\chi_{yy}= \chi_\perp$ and $\chi_{zz}=\chi_\parallel$ at $T=0$ as a function of {$c_2$}, taking constant $c_1=c_1(\hat{\mathbf{k}})$ and $c_2=c_2(\hat{\mathbf{k}})$ with normalization $c_1^2+c_2^2=1$, and using Eq.~\eqref{eq:suscNU_T0}. Note the reversal of hard/easy axes similar to the one shown in Fig.~\ref{fig:1}(b) for $O_h$.

The symmetry group $D_{6h}$ applies to UPt$_3$. Based on the presence of three superconducting phases in the $(H,T)$-phase diagram as well as the anisotropy of the upper critical field, a strong case was made for a chiral triplet order parameter in a two-dimensional IR, with $\mathbf{d_k}\parallel \hat z$ \cite{ChoiIdentificationoddparity1991, Saulsorderparametersuperconducting1994}. This corresponds to the pairing states indicated in the Table with $c_1=0$, making them unitary. The weak temperature dependence of the Knight shift reported for $\mathbf{H}\perp {\hat x}$~\cite{TouNonunitarySpinTriplet1998} is not in line with this scenario, nor with the nonunitary ones with $c_1\neq 0$. Alternative scenarios rely on the assumption of weak SOC and the rotation of $\mathbf{d_k}$ with $\mathbf{H}$ \cite{OhmiNonUnitaryTriplet1996}. Thus the question of the order parameter remains unsettled.

{\it Tetragonal symmetry.--} From the Table, we see that the nonunitary phase that may occur in $D_{4h}$ bears ressemblance with one of the possible nonunitary phases in $D_{6h}$. Recent Knight shift measurements for $\mathbf{H}$ in the basal plane of the tetragonal Sr$_2$RuO$_4$ crystal revealed a suppression that was not observed in earlier experiments~\cite{PustogowConstraintssuperconductingorder2019, IshidaReduction$17$OKnight2020,PetschReductionSpinSusceptibility2020}. They ruled out the long-discussed chiral (unitary) triplet phase, corresponding to the one that appears in the Table with $c_1=0$, and lead to suspect a singlet (rather than triplet) phase. Alternative nonunitary triplet scenarios would rely on $c_1\ll c_2$, as one expects the $\hat k_z$-dependence to be suppressed because of the layered (quasi two-dimensional) crystal structure of Sr$_2$RuO$_4$. That constraint would be relaxed in a three-dimensional scenario \cite{RoeisingSuperconductingorder$mathrmSr_2mathrmRuO_4$2019}, with the suppression of the spin susceptibility in the basal plane at $c_1\neq0$, which is illustrated in Fig.~\ref{fig:2}(b). Note also the multiband character of Sr$_2$RuO$_4$ \cite{KuhnAnisotropymultibandsuperconductivity2017}, which is not taken into account in our theory.

{\it Orthorhombic symmetry.--} All IRs are one-dimensional in the group $D_{2h}$ that applies to UTe$_2$, in which superconductivity was recently discovered \cite{RanNearlyferromagneticspin2019,AokiUnconventionalSuperconductivityUTe22022}. Thus scenarios of nonunitary superconductivity that have been proposed to interpret signatures of time-reversal symmetry breaking in the Kerr effect \cite{HayesMulticomponentsuperconductingorder2021} require that two IRs have accidentally close $T_c$s. Indeed, a recent microscopic study found evidence of such a near-degeneracy~\cite{IshizukaPeriodicAndersonmodel2021}. Various proposals result in $\mathbf{d_k}\propto (0,1,i\delta )$ with $\delta$ real and, thus, $\mathbf{q_k}\propto (1,0,0)$ \cite{AokiUnconventionalSuperconductivityUTe22022}. Our theory can be easily extended to almost degenerate IRs. In particular, the expression for the spin susceptibility, Eq.~\eqref{eq:chi_NU}, keeps its form, while the gap equations have to be generalized. Our results for the proposed form of $\mathbf{d_k}$ are consistent with the observed suppression of $\chi_{yy}$ and $\chi_{zz}$ below $T_c$ \cite{NakamineAnisotropicresponsespin2021}, also predicted in Ref.~\cite{HiranumaParamagneticEffectsj2021}, while $\chi_{xx}$ remains normal \cite{FujibayashiSuperconductingOrderParameter2022}.

{\it Conclusion.--}
Even though the spin susceptibility plays an important role in identifying the order parameter symmetry, it had not been calculated for general nonunitary phases. Here we filled this gap and analyzed various experimental results in the light of our findings. It paves the way for future studies on nonunitary superconductivity, including the role of multiple bands \cite{RamiresNonunitarysuperconductivitycomplex2022}, impurities, ferromagnetism in the normal state~\cite{MineevSuperconductivityuraniumferromagnets2017}, as well as finite field effects, in particular the paramagnetic limit of triplet superconductivity.

\begin{acknowledgments}
We thank J.-P. Brison and K. Hasselbach for useful discussions. We acknowledge the support of France 2030 ANR QuantForm-UGA and ANR-21-CE30-0035 (TRIPRES).
\end{acknowledgments}

\bibliography{Bibliography}
\bibliographystyle{apsrev4-1}

\end{document}